\documentclass[]{article}
\usepackage{graphicx}

\usepackage{hyperref}
\usepackage{url}
\usepackage{subcaption}
\usepackage{times}  
\usepackage{helvet} 
\usepackage{courier}  
\usepackage{graphicx} 
\usepackage{natbib}  
\usepackage{caption} 
\usepackage{graphicx}
\usepackage{xspace}
\usepackage{xcolor}

\usepackage{amsmath}
\usepackage{float}
\usepackage{hyperref}
 \usepackage{mathptmx, amsmath, amssymb, amsfonts, amsthm, mathptmx, enumerate, color}
 \usepackage{graphicx}
\usepackage{sidecap}
\usepackage{booktabs}
\usepackage{siunitx}
\usepackage{algpseudocode}

\newtheorem{theorem}{Theorem}

\newtheorem{lemma}[theorem]{Lemma}

\title{Generative Model Adversarial Training for Deep Compressed Sensing
}
\author{Ashkan Esmaeili
\thanks{Ashkan Esmaeili is with the Electrical and Computer Engineering Department, University of Central Florida (email: ashkan.esmaeili@ucf.edu).}
}

\begin{document}

\maketitle
\begin{abstract}
~~Deep compressed sensing assumes the data has sparse representation in a latent space, i.e., it is intrinsically of low-dimension. The original data is assumed to be mapped from a low-dimensional space through a low-to-high-dimensional generator. In this work, we propound how to design such a low-to-high dimensional deep learning-based generator  suiting for compressed sensing, while satisfying robustness to universal adversarial perturbations in the latent domain. We also justify why the noise is considered in the latent space.
The work is also buttressed with theoretical analysis on the robustness of the trained generator to adversarial perturbations. Experiments on real-world datasets are provided to substantiate the efficacy of the proposed \emph{generative model adversarial training for deep compressed sensing.}
\end{abstract}


\section{Introduction and Related Work}

Classical compressed sensing (CS) \cite{candes2008introduction} is built upon mapping a class of high-dimensional data into a low-dimensional one with the capability to reconstruct the high-dimensional data.
Mathematically, classical CS can be formalized as
\begin{equation}
   \mathbf{y}=\boldsymbol{\Phi}\mathbf{x},
\end{equation}
where $\mathbf{x} \in \mathbb{R}^{n}$ is the high-dimensional signal, and $\mathbf{y} \in \mathbb{R}^{m}$ is the sensed (compressed) signal (of lower dimension $m<<n$). 

 The inverse problem of \emph{low-to-high} dimensional reconstruction (retrieving $\mathbf{x}$ from observations $\mathbf{y}$) is not a uniquely doable task unless certain \emph{low-complexity} assumptions govern the \emph{data} distribution along necessary conditions on \emph{mapping/sensing} methods \cite{candes2008introduction}. Classical CS assumes the high-dimensional data $\mathbf{x}$ is sparse (as its low-complexity assumption), i. e., $\Vert \mathbf{x} \Vert_0 \ll n$ or has a sparse representation is some other domain. Moreover, if certain constraints hold for the sensing matrix (the restricted isometry property (RIP) for instance \cite{candes2011compressed}), unique reconstruction from low-dimensional space is guaranteed. 
 
 A vast and saturated literature on reconstruction guarantees and algorithms ($\ell_1$ programming as the most prominent one) has been developed in the past decades \cite{candes2008introduction, eldar2012compressed, esmaeili2016iterative}. Detailed review of CS literature is beyond the scope of this work  the proposed work centers on deep compressed sensing (DCS) models to be elaborated hereunder.

Ever-increasing works in deep learning establish the supremacy and robustness of generative models over discriminative ones \cite{goodfellow2016deep}. With this in mind, the overarching goal in DCS is to employ generative models which are expected to yield more robust and meritorious DCS models compared to the classical CS models. 
The shortcoming in the classical CS model is narrowing down the \emph{information-theoretical} low-dimensional input space into a limited class of the so-called $k$-sparse signals. Nevertheless, high-dimensional data can be generated from a low-dimensional latent space, where the low-complexity structure is implicitly captured as the embedded data lie on a low-dimensional latent space. This model is expected to be more robust in introducing sparse signals derived from  $\mathbb{R}^k \rightarrow \mathbb{R}^n$ mappings, since more diversity underlies such mappings leveraging deep learning generative models compared to the classical CS $k$-sparse signals. Moreover, it is expected that more robustness to adversarial perturbations is achieved in DCS thanks to the generative model power.
\begin{figure}
    \centering
    \includegraphics[width=6in, height=4.13in]{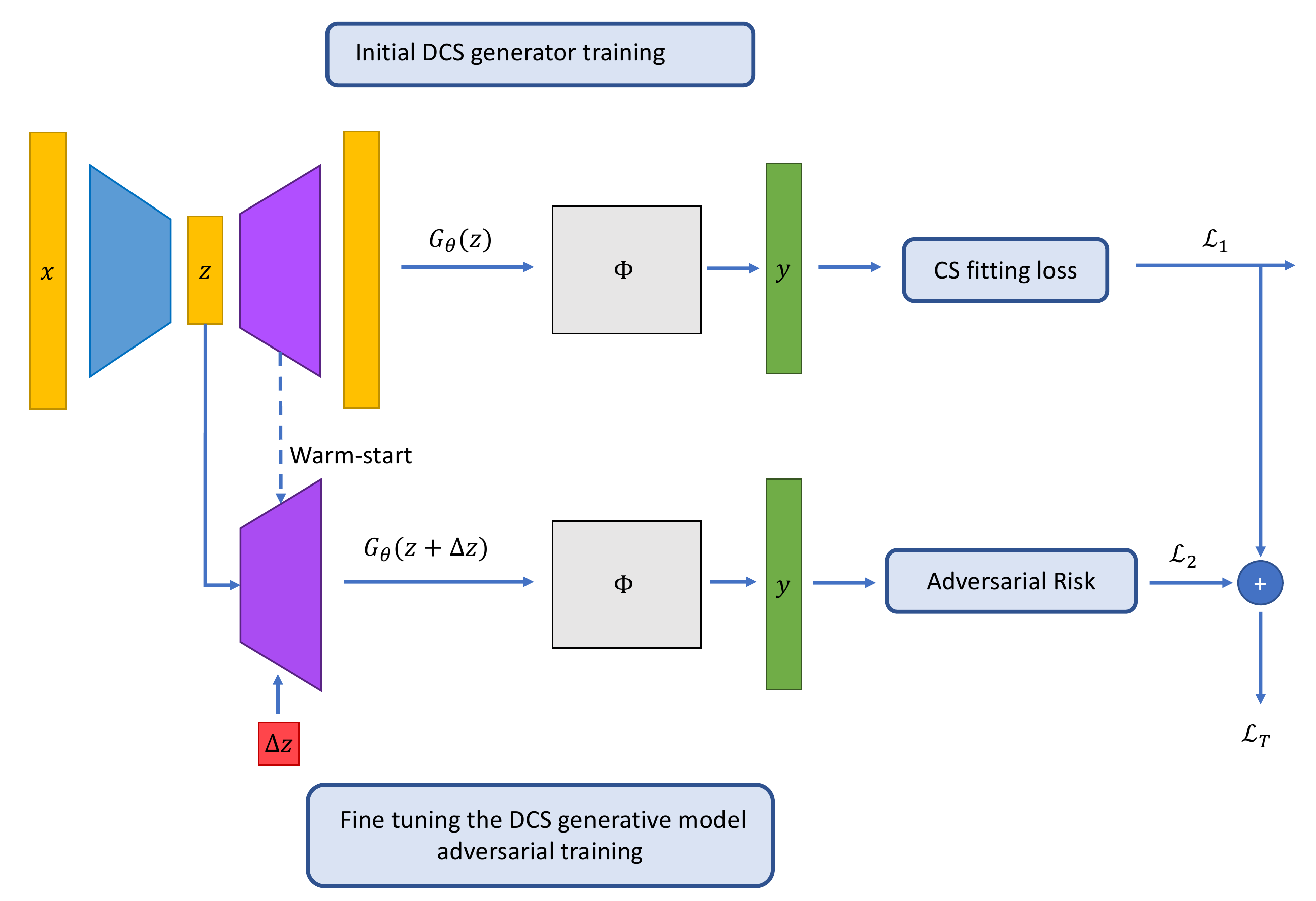}
    \caption{\footnotesize{The schematic representation of the deep compressed sensing adversarial training procedure set forth in the paper.}} 
    \label{fig:main}
\end{figure}

After this prelude, the mathematical representation of the DCS is given as $\boldsymbol{y}=\boldsymbol{\Phi}\boldsymbol{G}_{\theta}(\boldsymbol{z})$,
where $\boldsymbol{G}_{\theta}$ is the deep learning based $\mathbb{R}^k \rightarrow \mathbb{R}^n$ generator.
One can find utilization of generative models in DCS and how they generalize beyond classical CS models in \cite{wu2019deep, mardani2017deep, bora2017compressed,mardani2018deep,sun2020learning, van2018compressed}. 
There exists certain works which address the robust deep compressed sensing for defending against adversarial attack. The most important and relevant works are done by Jalal et. al. \cite{jalal2017robust} and \cite{jalal2020robust} in robust DCS adversarial training (DCSAT).
There are three main differences which distinguish our proposed work from the line of thought in the aforementioned: \textbf{1}- The adversarial risk in our setting is considered on the sensed domain rather than the high-dimensional signal (generator's output). \textbf{2-}
    In our work, we consider controlling the maximal perturbation only centering on real samples (images) to be sensed rather than sweeping for maximal deviation in the generators output space which is both a highly non-convex and high dimensional problem with exhaustive search space. Additionally, it exerts harsher design constraints on the generator compared to the former scenario.
    \textbf{3-} In \cite{jalal2017robust}, the generator is pre-trained and the adversarial perturbation affects the final model through the generator output. Contrarily, in our case, we assume certain latent representations (sparse domain) are available and translate the input noise to the latent space noise. The reason we consider the noise in the latent space in our model is two-folded: 1- Many practical systems such as communication systems assume that the signal which is encoded, i.e., the  input signal to the channel is directly affected by the additive noise. 2- The adversarial signal can be considered as non-robust features in the latent domain. It has been shown (as in   
    \cite{moosavi2017universal, akhtar2018defense}) that the perturbations in the latent space can be universal generalizable to a wide range of data. 

 Our method adapts and train the generator such that the robustness is realized for perturbations in the latent space.  
Further elaborations are provided through the paper body.
Several applications rely on supremacy of DCS models including medical resonance imaging (MRI) \cite{mardani2017deep, li2019segan, quan2018compressed, yi2019generative, jiang2019accelerating, qiusheng2020compressed, lee2017deep} and wireless neural recording \cite{sun2016deep}.
\vspace{-3mm}
\section{Deep Compressed Sensing Adversarial Training}\label{DCSAT}
In this section, we propose our DCSAT procedure as depicted in Fig. \ref{fig:main}.
The desired problem to train the generator equipped against adversarial attack can be cast as \footnote{
Each signal in the original domain $\boldsymbol{x}$, is characterized by a representation in the latent domain $\boldsymbol{z}$ as $\boldsymbol{x}=\boldsymbol{G_{\theta}}(\boldsymbol{z})$.
Hence, the joint distribution $p(\boldsymbol{x,y})=f_{\theta}(p(\boldsymbol{z,y}))$ can be expressed through a 1-1 mapping relating it to the joint distribution between sensed signals and the latent representation samples.},
\begin{align}\label{eq:adv}
\vspace{-2mm}
    \hat{\theta}(\lambda)=\underset{\theta}{\rm{argmin}}~ \Bigg [ \mathbb{E}_{(\boldsymbol{z,y}) \sim p(\boldsymbol{z,y})}\Big\{  \underbrace{\underset{\|\Delta z \|_2 \leq \epsilon}{\max}~~ \|\boldsymbol{y}-\boldsymbol{\Phi}\boldsymbol{G}_{\theta}(\boldsymbol{z+\Delta z}) \|_2^2}_{\textbf{Adversarial~risk}}
   + \lambda \underbrace{ \|\boldsymbol{y}-\boldsymbol{\Phi}\boldsymbol{G}_{\theta}(\boldsymbol{z})\|_2^2 \Big\}}_{\textbf{Fitting loss}}  \Bigg ]
\end{align}

The overall optimization \eqref{eq:adv} is carried out over the generator parameters $\theta$ so that two objectives (as characterized with braces) are met simultaneously, \textbf{1}- Nullifying the adversarial risk term which denotes the maximum deviation resulted from the adversarial attack $\boldsymbol{\Delta z}$ in the latent space. 
\textbf{2}- The second term is the CS fitting loss which aims to satisfy the DCS equation $\boldsymbol{y}=\boldsymbol{\Phi}\boldsymbol{G}_{\theta}(\boldsymbol{z})$. 
It worth noting (as shown later) that the first term acts like a regularization on the generator parameters and prevents overfitting resulted from the second term. In addition, the coefficient $\lambda$ establishes a trade-off between the induced regularization effect and the fitting term precision. One can consider the perturbation in the original domain which translates to some $\Delta z$ in the latent space.
\vspace{-2mm}
\section{Proposed Method}\label{sec:proposed}
\vspace{-1mm}
In this section, we expand on the DCSAT procedure.
The first order approximation for the generator $\boldsymbol{G}_{\theta}$ 
can be written as:
\vspace{-2mm}
\begin{align}\label{eq:4}
\boldsymbol{G}_{\theta}(\boldsymbol{z+\Delta z}) = \boldsymbol{G}_{\theta}(\boldsymbol{z})-\mathcal{J}_{\boldsymbol{G}}(\boldsymbol{z;\theta})\Delta \boldsymbol{z}
\end{align}
, where $\mathcal{J}_{\boldsymbol{G}}(\boldsymbol{z;\theta})$ is the Jacobian matrix of the generator function w.r.t $\theta$. Using the first order approximation, we simplify the adversarial risk so as to make it explicitly interpretable. Next, our proposed algorithm is built upon this simplification.

Let $\hat{\boldsymbol{y}}=\boldsymbol{y}-\boldsymbol{\Phi}\boldsymbol{G}_{\theta}(\boldsymbol{z})$ and $\boldsymbol{\Phi}\mathcal{J}_{\boldsymbol{G}}(\boldsymbol{z;\theta})$ denote $\boldsymbol{P}$.
As observed in P \eqref{eq:adv}, the adversarial risk loss value is obtained from a constrained optimization on $\Delta \boldsymbol{z}$ as the adversarial perturbation norm is limited. This leads to a quadratically constrained quadratic programming (QCQP) as follows:
\begin{align}\label{QCQP}
& \underset{\Delta \boldsymbol{z}}{\min}~~ -\Delta\boldsymbol{z}^T \boldsymbol{P}^T\boldsymbol{P}\Delta \boldsymbol{z}+2\hat{\boldsymbol{y}}^T\boldsymbol{P} \Delta \boldsymbol{z}-\|\hat{\boldsymbol{y}}\|_2^2  \nonumber \\
& \rm{subject~to}~~~~ \| \Delta \boldsymbol{z} \|_2 \leq \epsilon 
\end{align}
P \eqref{QCQP} is a trust-region problem and has a solution if and only if $\|\boldsymbol{z}\| \leq \epsilon$ and there exists $\mu \geq 0$ such that:
\begin{align}
& \textbf{1.}~~(-\boldsymbol{P}^T \boldsymbol{P}+\mu\boldsymbol{I}) \Delta \boldsymbol{z} = -\boldsymbol{P}^T \hat{\boldsymbol{y}}\\
& \textbf{2.}~~ (-\boldsymbol{P}^T \boldsymbol{P}+\mu\boldsymbol{I}) \succeq \bf{0} \label{PSD}\\
& \textbf{3.}~~ \mu(\epsilon-\|\Delta \boldsymbol{z}\|_2)=0 \label{eqcons}
\end{align}
It follows from \eqref{PSD}  that $\mu \geq \rm{eigmax(\boldsymbol{P}^T\boldsymbol{P})}$. Hence, \eqref{eqcons} leads to $\| \Delta\boldsymbol{z}\|_2=\epsilon$. Also, from (\textbf{5}) we have:
\begin{align}\label{sol:exp}
    \Delta \boldsymbol{z}=-(\mu\boldsymbol{I}-\boldsymbol{P}^T \boldsymbol{P})^{-1}\boldsymbol{P}^T \hat{\boldsymbol{y}}
\end{align}
Let $\boldsymbol{P}=\boldsymbol{U}\Sigma \boldsymbol{V}^T$ denote the singular value decomposition (SVD) for $\boldsymbol{P}$.
Using the SVD representation in Eq. \ref{sol:exp}, by expansion we obtain:
$
\Delta \boldsymbol{z}=-\boldsymbol{V}\rm{diag} \bigg \{ \frac{\Sigma_{ii}}{\mu-\Sigma_{ii}^2} \bigg\}\boldsymbol{U}^T\hat{\boldsymbol{y}}
$, from which the norm constraint equality on $\Delta \boldsymbol{z}$ follows as:
\begin{equation}\label{eq:norm}
    \big \langle \rm{diag} \bigg \{ \frac{\Sigma_{ii}}{\mu-\Lambda_{ii}} \bigg\},\boldsymbol{U}^T\hat{\boldsymbol{y}} \big \rangle ^2=
    \hat{\boldsymbol{y}}^T\boldsymbol{U}\rm{diag}\bigg\{\frac{\Lambda_{ii}}{(\mu-\Lambda_{ii})^2}\bigg\} \boldsymbol{U}^{T}\hat{\boldsymbol{y}}=\epsilon^2 
\end{equation}
\vspace{-5mm}
\subsection{Explicit Upper Bound for Adversarial Risk}
Utilizing the Cauchy-Schwartz inequality, the matrix operator norm inequality, and the energy constraint for $\Delta \boldsymbol{z}$, the adversarial risk can be upper bounded as: 
\begin{align}
   \Delta\boldsymbol{z}^T \boldsymbol{P}^T\boldsymbol{P}\Delta \boldsymbol{z}-2\hat{\boldsymbol{y}}^T\boldsymbol{P} \Delta \boldsymbol{z}+\|\hat{\boldsymbol{y}}\|_2^2 \leq 
   \|\hat{\boldsymbol{y}}\|_2^2+2\|\hat{\boldsymbol{y}}\| \|\boldsymbol{P}\|_{op}\|\epsilon\|+\|\boldsymbol{P}\|_{op}^2\|\epsilon\|_2^2
\end{align}
Using the fact that $|ab| \leq \frac{a^2+b^2}{2}$, the latter can be upper bounded with  
$
 2\|\hat{\boldsymbol{y}}\|_2^2+2\|\boldsymbol{P}\|_{op}^2 \epsilon^2$.
The upper bound is used in order to be make the adversarial risk explicitly expressible in terms of the network parameters and as a result, derive simpler algorithm for evaluating back-propagation gradient flows in training the generator. Although this is an upper bound, we establish in the following section that it is not a loose upper bound, i. e., minimizer of the upper bound also pushes down the proposed adversarial risk in P \eqref{eq:adv}.
Substituting the upper bound instead of the original adversarial risk, the dependency of the optimization problem on $\Delta \boldsymbol{z}$ is dropped, and
the resulted function is explicitly parameterized only on the generator parameters $\theta$ as:
\vspace{-3mm}
\footnote{
Working with training samples, the expectation in optimization \eqref{eq:app} is substituted with summation over training samples, i. e., pairs of $(\boldsymbol{z_i},\boldsymbol{y_i})$. 
  \begin{align}\label{p:exp}
  \vspace{-3mm}
   \hat{\theta}(\lambda)=\underset{\theta}{\rm{argmin}}~  \sum_{i=1}^n \|\boldsymbol{\Phi} \mathcal{J}_G(\boldsymbol{z_i};\theta)\|_{op}^2\epsilon^2 +
   \frac{\lambda+1}{2} \|\boldsymbol{y_i}-\boldsymbol{\Phi}\boldsymbol{G}_{\theta}(\boldsymbol{z_i})\|_2^2 
\end{align}}
\begin{align}\label{eq:app}
\vspace{-3mm}
   \hat{\theta}(\lambda)=\underset{\theta}{\rm{argmin}}~ \Bigg \{ \mathbb{E}_{(\boldsymbol{z,y}) \sim p(\boldsymbol{z,y})}\Big\{\|\boldsymbol{\Phi} \mathcal{J}_G(\boldsymbol{z};\theta)\|_{op}^2\epsilon^2 
   + \frac{\lambda+1}{2} \|\boldsymbol{y}-\boldsymbol{\Phi}\boldsymbol{G}_{\theta}(\boldsymbol{z})\|_2^2 \Big\}  \Bigg \} \vspace{-2mm}
\end{align} 
In training with back propagation, obtaining the gradient flow for the fitting loss is straightforward and can be evaluated as  $\mathcal{J}_{\boldsymbol{G}}^T(\theta)\boldsymbol{\Phi}^T(\boldsymbol{y}_i-\boldsymbol{\Phi}\boldsymbol{G}_{\theta}(\boldsymbol{z}_i))$. Yet, the approximation of the adversarial risk which contains the Jacobian $\mathcal{J}_{\boldsymbol{G}}(\boldsymbol{z}_i;\theta)$ must be explicitly expressed based on the network parameters which is considered in the following lemma \cite{zhang2019recurjac}. 
\begin{lemma}
Assume the generator $\boldsymbol{G}_{\theta}$ is an $H$-layer neural network. The Jacobian $\mathcal{J}_{\boldsymbol{G}}(\theta)$ can be written as:
\begin{equation}
    \vspace{-3mm}
    \boldsymbol{W}^H\boldsymbol{D}^H...\boldsymbol{W}^1\boldsymbol{D}^1, 
\end{equation}
where $\boldsymbol{D}^i$s are diagonal matrices containing the derivatives of the activation functions and $\boldsymbol{W}^i$s are the dense layer matrices.
\end{lemma}\label{lem:jac}
In order to minimize the operator norm of the product $\boldsymbol{P}=\boldsymbol{\Phi}\mathcal{J}(G_\theta;\boldsymbol{z})$, one can bound the operator norms of the layers $\boldsymbol{W}^l$ which constitute the Jacobian as specified in lemma \ref{lem:jac}. Owing to the Cauchy-Schwartz inequality for the product of matrices, the overall operator norm will be also bounded and regularized. We implement this by exerting the loss $\|\mathbf{I}-\boldsymbol{W}^T\boldsymbol{W}\|_F^2$, whose gradient is explicitly derivable, in place of the Jacobian operator norm term in P \eqref{eq:app}. Such regularization perform Lipschitz projection of the $\boldsymbol{W}$ layers. In DCS setting, the sensing matrix can be considered as the final dense layer. However, it is constant and not trainable. Accordingly, Lipschitz regularization cannot be applied on $\Phi$.
Instead, we project as much energy of $\boldsymbol{W}^H$ (the last layer) to the null-space of the sensing matrix $\boldsymbol{\Phi}$ by regularizing $\|\boldsymbol{\Phi} \boldsymbol{W}\|_{op}$. It is worth noting that although such regularizations help mitigating the adversarial attack effect, they also limit the learning and exploration capability of the model. A compromise between the CS fitting loss and the adversarial risk through tuning $\lambda$ determines the extent to which the regularizations are employed (The ensemble parameters $\boldsymbol{W}^l, \boldsymbol{D}$ are denoted in $\theta$).

\vspace{-3mm}
\subsection{Why Do Upper Bound Optimal Parameters Also Apply to the Original Adversarial Risk?}
Substituting the closed form $\Delta \boldsymbol{z}$ in objective function \eqref{QCQP}, we have: 
\begin{equation}
    \mathcal{L}= \hat{\boldsymbol{y}}^T\boldsymbol{U}\rm{diag}\bigg\{1+\frac{2\Lambda_{ii}}{\mu-\Lambda_{ii}}+\frac{\Lambda_{ii}^2}{(\mu-\Lambda_{ii})^2}\bigg\} \boldsymbol{U}^T\hat{\boldsymbol{y}}. \label{eq:mu}
    \end{equation}
Now, we adjudicate how minimizing the approximated upper bound instead of the proposed adversarial risk, and using the achieved parameters also decreases the adversarial risk function in P \eqref{eq:adv}.
If the operator norms of $\boldsymbol{W}^l$s are regularized, the exploration of the learning function reduces which ends up with higher fitting loss $\hat{\boldsymbol{y}}$. 
The benefit of regularization towards mitigating the adversarial risk during the training procedure must outweigh the model learning degradation (i. e., increase in the fitting loss $\hat{\boldsymbol{y}}$) to yield an acceptable adversarial training. 
If the coefficient $\lambda$ is set to a large value in P \eqref{eq:adv}, only slight increase in $\hat{\boldsymbol{y}}$ is tolerable as a result of regularization.
Otherwise, the large coefficient $\lambda$ leads to a large gap $\lambda \Delta {\hat{y}}$ that can not be compensated through the positive effect of regularization applied through adversarial risk. Thus, the change in $\hat{\boldsymbol{y}}$ becomes negligible with a large choice for $\lambda$ ( regularization coefficient for fine-tuning $\boldsymbol{G}_{\theta})$. Also, the basis span in SVD for $\mathcal{J}_{\boldsymbol{G}}(\theta)$ and as a result in $\boldsymbol{P}$ is assumed not to rotate as the proposed Lipschitz projection simply regularizes the singular values and does not change the exploring subspace found for $\boldsymbol{W}$ layers meaning that $\boldsymbol{U}$ and therefore $\boldsymbol{U}^T\hat{\boldsymbol{y}}$ are invariant. 
With this in mind, taking a look back into the norm constraint Eq. \eqref{eq:norm}, $\Lambda_{ii}$s are reduced after regularization. Unless $\mu$ is reduced appropriately, the equality would not be maintained.
Now, consider the loss in Eq. \eqref{eq:mu}. 
The change in the first term was forced to be negligible by proper choice for $\lambda$. We write the second and third diagonal terms as $2\frac{\Sigma_{ii}}{(\mu-\Lambda_{ii})}\Sigma_{ii}$ and $\frac{\Lambda_{ii}}{(\mu-\Lambda_{ii})^2}\Lambda_{ii}$ , respectively. In both fractions, the first term is invariant, and the second terms are eigenvalues which are regularized (reduced). Hence, the whole term shrinks leading to shrinkage of the original adversarial risk.
All in all, addressing Lipschitz property for the surrogate optimization leads to smaller adversarial risk in the main proposed optimization and the upper bound surrogate is not a loose upper bound whose minimizer leaves no change in the original adversarial risk. Heuristically, $\mu$ decreases proportional to $\Lambda_{ii}$. Hence, the $\mu^2$ appearing in the nominator of the original adversarial risk decreases quadratically reducing the loss function more. 
\vspace{-3mm}
\section{Numerical Experiments}\label{sec:sim}
Images and in general large data arrays may be sensed only partially (a.k.a) missing data due to limitations in imaging devices (include example for MRI) or corrupt measurements.

In this work, we assume the sensing matrix is a random sampler which partially masks the images. This resembles the missing data and inference using a model-based structure on data. Rather than classical low-rank assumptions (as in \cite{azghani2019missing,esmaeili2018transduction,esmaeili2019novel}) we use the intrinsic low-dimensionality of latent space mapped with a deep learning model.

The objective is to reconstruct the original samples from the randomly sampled image. 
With such sensing method, which happens to be common in many practical applications, the attacker can invest only on parts of signal which happen to have overlap with the sensing matrix mask and hence, effectively expend the perturbation energy on the mask. Hence, a sampling mask makes the defensive training of the generator more difficult.
\subsection{Applying Adversarial Attack}
Next, we investigate the efficacy of our generative model DCSAT method in mitigating adversarial attack for certain simulation scenarios.
In order to apply the adversarial attack, we use omni-directional $\Delta \mathbf{z}$ directions around the initial latent sample and make consecutive queries to pick the maximum deviation in the sensed generator's output as the adversarial attack.
This is an empirical approach to find the adversarial attack. 

\subsection{MNIST Dataset} 
In this experiment, we have considered a generator consisting of two consecutive dense layers, mapping from the latent space with size $30$ to $100$, and from $100$ to $784$ (output image size). Next, the output is sensed with a random sampling matrix. The latent space representation of the MNIST data is obtained through a compressing decoder which maps data through two dense layers from $784$ to $100$ and $100$ to $30$, respectively. The encoder and decoder can be trained using an auto-encoder. After training the auto-encoder, the generator part is removed from the auto-encoder in order to be fine-tuned for DCS adversarial training. It can benefit from the warm-start obtained from training the auto-encoder. 
The latent representations are also stored to be utilized further as the generator input for generating high dimensional samples.
\begin{table}[]
    \centering
\begin{tabular}{|l|l|c|c|c|}
\hline
Method & ~~~ $\lambda/ SR$ & Adv. risk & Fit. loss    & total loss\\ \hline \hline
DCS          & ~~ \textemdash / 0.8  &    5.5948    & \textbf{4.0290} & 9.6238\\ 
\hline
DCSAT         & 200,000 / 0.8  &  5.5114   &   4.0370   & 9.5484 \\ 
\hline
DCSAT         & 20,000 / 0.8 &  5.3726   &  4.1051 & \textbf{9.4777}\\ 
\hline
DCSAT         & 2,000 / 0.8 &  \textbf{5.2547}    & 4.4278  & 9.6825\\ 
\hline
\hline
DCS          & ~~ \textemdash / 0.6 &   6.2214   & 5.2129  & 11.4343 \\ 

\hline
DCSAT         & 200,000 / 0.6 &    6.1611    &  \textbf{5.1508} & 11.3119\\ 

\hline
DCSAT         & 20,000 / 0.6 &     6.0561  & 5.2535  & \textbf{11.3096}\\ 
\hline

DCSAT         & 2,000 / 0.6 &     \textbf{5.8972}  & 5.4452 & 11.3424\\
\hline 

\end{tabular}
\vspace{-1mm}
    \caption{DCSAT ablation study on the MNIST test dataset.}
    \label{tab:mnist}
\end{table}
In the final step, the warm-start generator ought to be fine-tuned to both hold in the CS fitting loss as well as maintaining the discussed Lipschitz property through explicit adversarial regularizations.
In the final step, we attach the sensing matrix to the last dense layer of the generator and set its trainable option to false. We train the resulted generator to map the obtained latent representations to the sensed MNIST images.
\vspace{-3mm}
\subsection{CIFAR-10 Dataset}
\begin{table}[]
    \centering
\begin{tabular}{|l|l|c|c|c|}
\hline
Method & $ ~~~\lambda / SR$ & Adv. risk & Fit. loss & total loss    \\ \hline \hline
DCS          &~~ \textemdash  / 0.8  & 0.1375  &  \textbf{0.0184} & 0.1559\\ 
\hline
DCSAT         & 10,000 / 0.8  &   \textbf{0.0839}    & 0.0192 & \textbf{0.1029}\\ 
\hline
DCSAT         & 1000 / 0.8 &    0.1292   & 0.0213  & 0.1505\\ 
\hline
DCSAT         & 100 / 0.8 &  fail  & 0.0227  & \textemdash\\
\hline
\hline
DCS          &~~ \textemdash  / 0.6 &  0.1327   & \textbf{0.0166}  &  0.1493 \\ 

\hline
DCSAT         & 10,000 / 0.6 &  \textbf{0.0793}    & 0.0176  & \textbf{0.0969}\\ 

\hline
DCSAT         & 1000 / 0.6 &    0.1202  & 0.0194 & 0.1396\\ 

\hline
DCSAT         & 100 / 0.6 & fail & 0.0235  & \textemdash\\
\hline
\end{tabular}
      \caption{DCSAT ablation study on the CIFAR-10 test dataset. Larger errors for MNIST are due to early stopping (fewer epochs for fine-tuning).}
    \label{tab:cif}
\end{table}

In this experiment, the encoder and the decoders consist of convolutional layers rather than simple dense layers to extract local features \cite{balntas2016learning}. The inputs are CIFAR-10 images of size $(32\times32\times3)$. The encoder consists of three Conv2D layers which are followed by batch normalization layers, relu activation layers, and $(2,2)$ MaxPooling2D layers. The Conv2D layers have $64$, $32$, and $16$ filters, respectively. The kernel size is considered as $(3,3)$, and the stride is set to $(1,1)$. The decoder builds up reversely utilizing UpSampling2D to compensate for the MaxPooling2D layers.
Similar to the mnist setting, an auto-encoder is trained to find the encoded latent representations. Next, the decoder part can be fine-tuned via the proposed DCSAT to robustify the generator against adversarial perturbations while maintaining desired DCS precision.

\subsection{Simulation Result Analysis}
In this section, we discuss the simulation results and provide examples of how adversarially training the generator 
\begin{figure}
\vspace{-3mm}
\centering
\subfloat[CIFAR-10 $\lambda=1000$]{\includegraphics[height=2in]{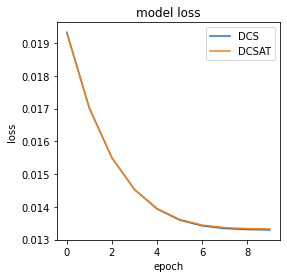}}
\subfloat[CIFAR-10 $\lambda=100$]{\includegraphics[height=2in]{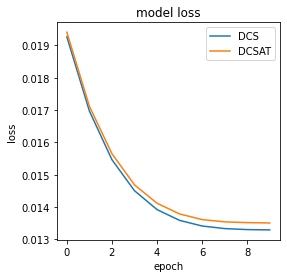}}
\subfloat[CIFAR-10 $\lambda=0.1$]{\includegraphics[height=2in]{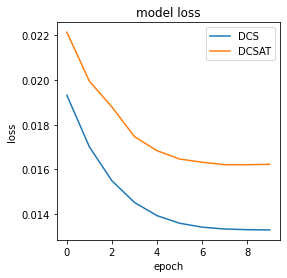}}\\
\subfloat[Cifar $\lambda=200,000$]{\includegraphics[height=2in]{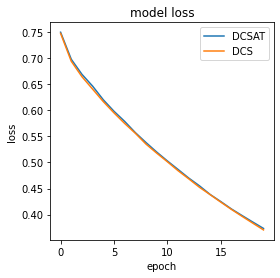}}
\subfloat[MNIST $\lambda=200,000$]{\includegraphics[height=2in]{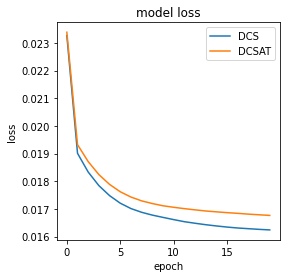}}
\subfloat[MNIST $\lambda=2000$]{\includegraphics[height=2in]{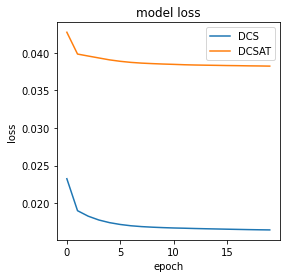}}
\caption{\footnotesize{Comparison of DCS vs. DCSAT training error for different regularization levels ($SR=80 \%$).}}\label{harze}
\end{figure}
In this section, we discuss the simulation results on test sets. Tables \ref{tab:mnist} and \ref{tab:cif} include an ablation study of two hyper parameters: 1- the adversarial/fitting loss trade-off ($\lambda$), and 2- two sampling rates (SR) ($80 \%$ and $60 \%$) for the sensing matrix $\Phi$.
The regularization parameter $\lambda$ compromises between the model robustness to perturbations and the learning capacity. More regularization induce learning bottleneck for the model. However, the  
As observed, regularization conversely affects fitting loss and best fitting loss values are related to DCS implementation where there is no adversarial training (except for one instance due to the overfitting effect). Contrarily, increasing $\lambda$, the adversarial risk  goes down except for ($\lambda < 1000$)  in CIFAR-10 experiment where the regularization bottleneck makes the model malfunction in learning. The optimal values are highlighted in bold. It is worth mentioning that the optimal design (minimal aggregate loss) is obtained from a hyper-parameter tuned DCSAT scenario and not the sole DCS which clarifies the supremacy of DCSAT over DCS. Fig.\ref{harze} shows the splitting behavior in training DCSAT and DCS models. 
\footnote{The experiments are available at \href{https://github.com/ashkanucf/DCSAT.git}{https://github.com/ashkanucf/DCSAT.git}.}

\section{Conclusion}
Deep compressed sensing is a more robust framework compared to classical CS. Adversarial attackers try to lower a system performance by triggering the signals in low-dimensional latent space resulting in significant variation of the sensed output. In this work, we have shown how to train a deep compressed sensing generator which is robust to the effect of universal perturbations triggered in the latent space as well as mapping data to the compressed sensing domain to suit for the compressed sensing loss. The work is buttressed with math analysis on how the applied method helps reduce the adversarial effect. Real-world compressed sensing experiments verify the efficacy of the proposed procedure in training the deep compressed sensing generator.  
\vspace{-2mm}
\bibliographystyle{plainnat}
\bibliography{main}

\end{document}